\documentclass[useAMS,usenatbib]{mnras}

\usepackage{graphicx}
\usepackage{epstopdf}
\usepackage{amsmath,esint}

\title[Magnetized Outflows from Stellar-Mass Black Holes]{On the Theoretical Framework of Magnetized Outflows from Stellar-Mass Black Holes and Related Observations}

\author[D. M. Christodoulou et al.]{D. M. Christodoulou,$^{1,2}$\thanks{E-mail: dimitris\_christodoulou@uml.edu} 
I. Contopoulos,$^{3,4}$\thanks{E-mail: icontop@academyofathens.gr} 
D. Kazanas,$^{5}$\thanks{E-mail: demos.kazanas@nasa.gov}
J. F. Steiner,$^{6}$\thanks{E-mail: jsteiner@mit.edu} \and
D. B. Papadopoulos,$^{7}$\thanks{E-mail: papadop@astro.auth.gr}
and S. G. T. Laycock$^{1,8}$\thanks{E-mail: silas\_laycock@uml.edu}
\\
$^{1}$Lowell Center for Space Science and Technology, University of Massachusetts Lowell, Lowell, MA, 01854, USA\\
$^{2}$Department of Mathematical Sciences, University of Massachusetts Lowell, Lowell, MA, 01854, USA\\
$^{3}$Research Center for Astronomy and Applied Mathematics, Academy of Athens, Athens 11527, Greece \\
$^{4}$National Research Nuclear University, Moscow 115409, Russia \\
$^{5}$NASA Goddard Space Flight Center, Laboratory for High-Energy Astrophysics, Code 663, Greenbelt, MD 20771, USA \\
$^{6}$MIT Kavli Institute for Astrophysics and Space Research, Cambridge, MA 02139, USA \\
$^{7}$Department of Physics, Aristotle University of Thessaloniki, Thessaloniki 54124, Greece \\
$^{8}$Department of Physics \& Applied Physics, University of Massachusetts Lowell, Lowell, MA, 01854, USA
}

\begin{document}

\pagerange{\pageref{firstpage}--\pageref{lastpage}} \pubyear{2016}

\maketitle

\label{firstpage}

\begin{abstract}
The spins of stellar-mass black holes (BHs) and the power outputs of their jets are measurable quantities. Unfortunately, the currently employed methods do not agree and the results are controversial. Two major issues concern the measurements of BH spin and beam (jet) power. The former issue can be resolved by future observations. But the latter issue can be resolved now, if we pay attention to what is expected from theoretical considerations. The question of whether a correlation has been found between the power outputs of few objects and the spins of their BHs is moot because BH beam power does not scale with the square of the spin of the BH. We show that the theoretical BH beam power is a strongly nonlinear function of spin that cannot be approximated by a quadratic relation, as is generally stated when the influence of the magnetic field is not accounted for in the \cite{bla77} model. The BH beam power of ballistic jets should scale a lot more steeply with BH spin irrespective of the magnetic field assumed to thread the horizon and the spin range considered. This behavior may already be visible in the analyses of radio observations by \cite{nar12} and \cite{rus13}. In agreement with previous studies, we also find that the power output that originates in the inner regions of the surrounding accretion disks is higher than that from the BHs and it cannot be ignored in investigations of continuous compact jets from these systems.
\end{abstract}





\section{Introduction}\label{intro}

In the past few years, a great deal of effort has been devoted to the study
of stellar-mass accreting black holes (BHs) that are members of binary systems 
with the intention to pinpoint the location of the innermost stable circular orbits
(ISCOs) of their accretion disks and to determine the state of rotation of these
compact objects. Such investigations are not easy to undertake as they
require high-resolution observations of the few known 
BH binaries followed by intensive relativistic modeling of the spectral features
that arise from the gas orbiting near and accreted by the BHs. Thorough
descriptions of the subject and the most accurate results to date
can be found in the reviews of \cite{rem06}, \cite{mil07}, \cite{mcc11,mcc14}, and
\cite{rey14}, 
as well as in the works of \cite{fen10}, \cite{mil11}, \cite{ste13,ste14}, \cite{reid14}, \cite{oro14}, \cite{wu16}, and \cite{che16}. 
In summary:
\begin{itemize}
\item[(a)] The continuum-fitting (CF) method is currently a leading method for determining 
the location of the ISCO of the orbiting gas and from it the BH spin parameter.
The CF method uses a model-dependent fit of the thermal 
continuum X-ray spectrum in low-luminosity systems and requires precise knowledge
of the physical state of the inner accretion disk, its inclination, the BH mass, and the 
distance to the system. Determinations of these parameters have been improving 
over the years and detailed model fits have produced accurate values of the location 
of the ISCO and the BH spin for a sample of 10 stellar-mass BHs 
\citep{mcc06,sha06,gou09,gou10,ste10,mcc11,gou11,ste14,che16}. \\
\item[(b)] The iron K$\alpha$-line (Fe K$\alpha$) method is an independent method that is
also used to determine the locations of the ISCOs in the accretion disks around
galactic as well as extragalactic BHs. This method uses a model-dependent fit of the
dynamical broadening of the Fe K$\alpha$ emission line that is excited in the 
inner edge of the accretion disk by external irradiation. The Fe K$\alpha$ method 
does not require knowledge of the BH mass, the disk inclination, or the distance to 
the system, and it has also produced accurate values of the location of the ISCO and 
the BH spin for a sample of about 10 stellar-mass BHs 
\citep{blu09,mil09,rei09,rei10,hie10,ste10,mil11,rus13,rey14}.
\end{itemize}

At present, seven BHs in X-ray binaries have been analyzed by both methods 
and the results disagree in the cases of 4U 1543-47 
and GRO J1655-40 \citep{rey14}; and likely in the case of GX 339-4
\citep{mil09,kol10} for which BH mass, distance, and disk inclination
are uncertain \citep{sha09,oze10}.
In this investigation,
we do not intend to examine the differences between the two methods;
on the contrary, we need a homogeneous data set of BH masses and spins
in order to investigate theoretically the electromagnetic output from such systems.
For this reason, we choose to use only data from the CF method which depends on accurate determinations of BH masses, values that we need for our study as well.
These data
are listed in Table~\ref{table1} and references are shown in the last column of the table.

\begin{table*}
\caption{Spin Parameters from the CF Method and Masses of Stellar-Mass Black Holes}
\label{table1}
\begin{tabular}{lllll}
\hline
No. & Object & $a_*$ & $M/M_\odot$ & References \\
\hline
1  & Cygnus X-1        & $>$0.99  		                & 14.8  ($\pm1.0$)    & 1, 2      \\
2  & GRS 1915+105  & $>$0.98                        & 12.4  ($+2.0,-1.8$)    & 3, 11    \\
3  & LMC X-1	            & 0.92 ($+0.05, -0.07$)    & 10.91 ($\pm1.54$)  & 4          \\
4  & M33 X-7	            & 0.84 ($\pm0.05$)         & 15.65 ($\pm1.45$)  & 5          \\
5  & 4U 1543-47	    & 0.80 ($\pm0.05$)         & 9.4   ($\pm1.0$)      & 6          \\
6  & GRO J1655-40   & 0.70 ($\pm0.05$)         & 6.30  ($\pm0.27$)   & 6          \\
7  & Nova Mus $'$91  & 0.63 ($\pm0.18$)         & 11.0  ($+2.1,-1.4$)   & 12, 13          \\
8  & XTE J1550-564   & 0.34 ($+0.20,-0.28$)   & 9.10  ($\pm0.61$)    & 7              \\
9  & LMC X-3	             & 0.25 ($\pm0.15$)        & 7.0  ($\pm0.6$)       & 8, 10, 14, 15   \\
10 & A0620-00	         & 0.12 ($\pm0.19$)         & 6.61  ($\pm0.25$)   & 9          \\
\hline
\end{tabular}
\\
\smallskip
{Ref. Key: 1--\cite{oro11}, 2--\cite{gou11}, 3--\cite{mcc06}, 4--\cite{gou09}, 5--\cite{liu08,liu10}, 6--\cite{sha06}, 7--\cite{ste10}, 8--\cite{dav06}, 9--\cite{gou10}, 10--\cite{fen10}, 11--\cite{reid14}, 12--\cite{wu16}, 13--\cite{che16}, 14--\cite{oro14}, 15--\cite{ste14}}
\end{table*}

Determinations of the ISCOs and the spins of BHs have consistently neglected
the effects of the magnetic field that may exist in the inner accretion disks
and in the accreted plasma. The rationale behind this assumption is that
torques due to the embedded magnetic field at the ISCO can only account for 
a small correction that lies well within the errors due to modeling of the X-ray 
continuum in low-luminosity sources ($\ell/\ell_{\rm Edd} < 0.3$, where 
$\ell_{\rm Edd}$ is the Eddington luminosity). 
As detailed by \cite{mcc11}, calculations that take magnetic 
torques into account do not produce systematic errors larger than the 
observational errors, even for disks with $\ell/\ell_{\rm Edd}\approx 0.35-0.5$.

However, the above efforts to quantify the influence of an embedded
magnetic field near the ISCO do not address the problem in its entirety because the power output of jet-like outflows from these
systems depends strongly on the magnetic field that can be supported in the region between the ISCO and the event horizon of the BH \citep{con12,con16}.
Furthermore, it is now understood that magnetic field can be generated {\it in situ} and grow linearly 
in the inner disks around BHs by the Poynting-Robertson Cosmic Battery (PRCB) 
\citep{ck98,con06,chr08},
and this mechanism appears to be at work in the vicinities of both 
supermassive BHs \citep{con09,chr16}
as well as stellar-mass BHs \citep{kyl12}.
The PRCB can operate efficiently in the inner accretion disks of stellar-mass
compact systems for as long as the accretion flows are advection-dominated (ADAF),
and it is capable of building a significant magnetic field of a single polarity in less than 1 day \citep{kyl12}.

Generally speaking, accretion disks around stellar-mass BHs have two ``magnetic" states available to them depending on whether their BHs were born with a very high or a low/moderate spin. 
Because of the extremely long timescales for substantially increasing the spin by accretion 
\citep{mcc11,gou11},
the various compact systems cannot cross over between 
these two states and the observations should then find them with their ISCOs 
either near the event horizon or near the nonrotating value \citep{con12}.
Thus, the result could be a segregation of spins into two broad groups.
The data shown in Table~\ref{table1} already appear to be in agreement 
with this hypothesis despite the markedly small number 
of objects involved; there exists only one object with spin $a_*$
(defined in \S~\ref{bhonly} below)
in the intermediate
region between $0.34$ and $0.70$. 
A similar gap in the $a_*$ interval of 0.3-0.75 is seen in the Fe~K$\alpha$ data as well \citep{fen10}. For these reasons, we are compelled to analyze the physical properties of the two groups of objects separately and not just as one uniform sample. 

In what follows, we describe the theoretical framework for the energetics of jet-like outflows from stellar-mass BH binaries. The expected power output from
such systems is analyzed in considerably more detail than previously done
\citep{liv99,mei99,mei01}.
Equipartition between the magnetic pressure and the ram pressure of inflowing matter provides an estimate of the magnetic
field that can be supported in the region between the ISCO and the BH horizon \citep{ck98}.
Then the poloidal magnetic flux and the beam (jet) power are calculated by using
the standard equations provided by classical theories of emission from the BH
and the inner accretion disk \citep{bla77,bla82,con94,liv99,mei99,mei01}. 
In \S~\ref{bhonly}, we analyze jet emission only from the vicinity of the BH.
In \S~\ref{total}, we add the contribution from a nonrelativistic disk to the results of \S~\ref{bhonly}. In \S~\ref{new}, we repeat the calculations 
using a new estimate of the maximum supported magnetic field inside the ISCO \citep{con16}. 
We conclude in \S~\ref{conc} with a summary and a discussion of our results.

\section{Electromagnetic Output from the Black Hole}\label{bhonly}

We determine the poloidal magnetic flux $\Psi_{BH}$ and the power
output $L_{BH}$ in the vicinity of a BH. Metric-system units are used
throughout (always appearing inside parentheses), unless stated otherwise. We adopt $R_I$ for the radius of the ISCO, 
\begin{equation}
R_S = \frac{2GM}{c^2} \ , 
\label{rsch}
\end{equation}
for the Schwarzschild radius, and the dimensionless ratio
\begin{equation}
x \equiv \frac{R_I}{R_S} \ .
\label{x_par}
\end{equation}
Here $M$ represents the BH mass, $c$ is the speed of light, and $G$ is the gravitational constant. 
The parameter $x$ is a nonlinear function\footnote{Eq.~(2.21) of \cite{bar72} provides the radius of the ``marginally stable orbit'' in geometric units as $r_{ms}/M \equiv f_*(a_*)$, where
$f_*(a_*)$ is a nonlinear function of the dimensionless spin parameter. For our purposes, $r_{ms}$ is identified with $R_I$ and $x$ is determined from the equation $x = R_I/R_S = f_*(a_*)/2$. We note that this equation does not take into account the more complicated model of \cite{con12} in which $x$ depends also on the
magnitude of the magnetic field itself.}
of the dimensionless spin parameter $a_*$  of the BH \citep{bar72}, where 
\begin{equation}
a_* \equiv \frac{cJ}{G M^2} \ ,
\label{a_star}
\end{equation}
and $J$ represents the angular momentum of the BH.

The calculations incorporate
the equipartition magnetic field $B_{eq}$ derived by \cite{ck98} 
(their eq.~[11]) for geometrically thick (ADAF) accretion disks 
\begin{equation}
B_{eq} = (9\times 10^{3}~{\rm T}) \ {\dot m}^{0.5} 
      \left(\frac{M}{M_\odot}\right)^{-0.5} x^{-1.25}\ ,
\label{ckb}
\end{equation}
where the mass of the BH is scaled in units of the solar mass $M_\odot$
and $\dot m$ is the mass accretion rate in units of its Eddington value.
This equation provides an estimate of the strongest magnetic field that may be supported throughout the region $R_H\leq R\leq R_I$ between the BH horizon and the ISCO.
The value of $B_{eq}$ depends on the location of the ISCO through the
$x^{-1.25}$ term, so the ISCO has a say (by as much as a factor of $\sim$10) on how much magnetic field will thread the BH horizon. The BH horizon radius 
$R_H$ is also
a nonlinear function of the spin \citep{bar72} and we can write \citep{rie11}
\begin{equation}
R_H = \frac{1}{2} \ R_S \ Q_*(a_*) \ ,
\label{hor}
\end{equation}
where
\begin{equation}
Q_*(a_*) \equiv 1 + \sqrt{1 - a_*^2} \ .
\label{qstar}
\end{equation}
For the case of no rotation ($a_*=0$), the horizon coincides with $R_S$, whereas in the extreme-rotation case with $a_*=1$, the horizon shrinks to $R_S/2$. We also need the rotation frequency on the horizon $\Omega_H$ at which inertial frames are dragged by the BH at $R_H$, and this is obtained from the equation \citep{rie11}
\begin{equation}
\Omega_H R_H = \frac{1}{2}\ c\ a_* \ .
\label{omegah}
\end{equation}
As a check, using eqs.~(\ref{rsch}) and~(\ref{hor}), we can cast eq.~(\ref{omegah})
to the form 
$\Omega_H = a_*/(2 M Q_*)$ in geometric units $c=G=1$,
as was defined by \cite{ste13}.
 
\begin{figure}
\begin{center}
    \leavevmode
      \includegraphics[trim=0 1.3cm 0 1cm, clip, angle=0,width=9 cm]{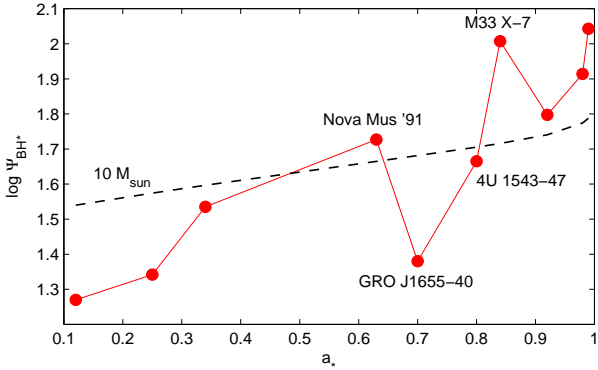}
\caption{Poloidal magnetic flux $\Psi_{BH*}$ from
eq.~(\ref{Psibh2}) vs. $a_*$ for the data listed in Table~\ref{table1}
and connected by straight line segments.
The dashed line shows a BH with a fixed mass of 10$M_\odot$.
In order of increasing $a_*$, the absolute errors in $\log\Psi_{BH*}$ are 
0.08, 0.09, 0.11, 0.17, 0.04, 0.08, 0.07, 0.12, 0.12, and 0.06.
\label{fig1}}
  \end{center}
  \vspace{-0.5cm}
\end{figure}

For the poloidal magnetic flux, we find that
\begin{eqnarray}
\Psi_{BH} & = & B_{eq}\left(\pi R_H^2\right)\nonumber \\
      & = & (6.2\times 10^{10}~{\rm Wb})\ {\dot m}^{0.5}\nonumber \\
      &    & \times \left(\frac{M}{M_\odot}\right)^{1.5} x^{-1.25} \ Q_*^2 \ ,
\label{Psibh1}
\end{eqnarray}
and we take the dimensionless flux $\Psi_{BH*}$ to be
\begin{equation}
\Psi_{BH*}\equiv \left(\frac{M}{M_\odot}\right)^{1.5} x^{-1.25}\  Q_*^2 \ ,
\label{Psibh2}
\end{equation}
in units of $(6.2\times 10^{10}~{\rm Wb}) {\dot m}^{0.5}$.
This quantity is not directly observable, but it is still useful to
examine its behavior in the few stellar-mass BHs that have so far been
studied extensively. For this purpose, we adopt an homogeneous data set of 10
BHs (Table~\ref{table1}) whose masses and spins have presently been determined as accurately as possible. The spins were all derived by the CF 
method.
Fig.~\ref{fig1} shows $\Psi_{BH*}$ from
eq.~(\ref{Psibh2}) vs. $a_*$ for the data listed in Table~\ref{table1}.
GRO J1655-40 has smaller mass than neighboring BHs in Fig.~\ref{fig1} and it appears to separate the fluxes into two groups,\footnote{There also exists another data set of spins derived by the Fe~K$\alpha$ method \citep{fen10,rus13,rey14}; these data do not disagree with our determinations from the CF data.
A similar segregation of objects into two groups is seen
in the Fe~K$\alpha$ data as well, but a different BH with a markedly small mass is responsible for the gap seen in magnetic fluxes. This BH is XTE J1650-500
with $M = (5\pm2.3)~M_\odot$ and $a_*=0.79\pm0.01$ \citep{oro04,mil09,oze10}.\label{ft2}}
one with increased fluxes at 
moderate $a_*$ values, and another with the highest fluxes at high
$a_*$ values.
It is hard to tell whether the effect is significant with so few data available.
Nevertheless, we are compelled to consider these two groups separately in the analysis
that follows.

 \begin{figure}
\begin{center}
    \leavevmode
      \includegraphics[trim=0 1.3cm 0 1cm, clip, angle=0,width=9 cm]{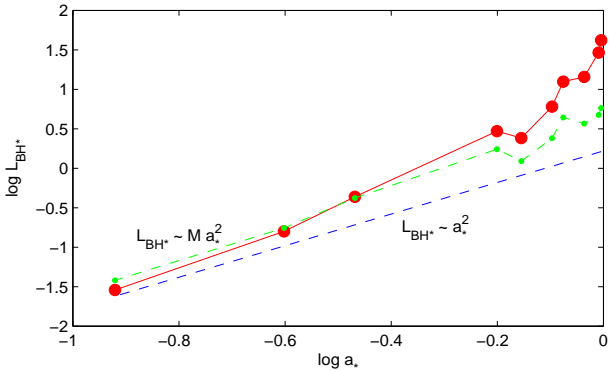}
\caption{Beam power $L_{BH*}$ from
eq.~(\ref{Lbh2}) vs. $a_*$ for the data listed in Table~\ref{table1}
and connected by straight line segments.
The dashed lines show a pure $a_*^2$ dependence (blue line)
and an $M a_*^2$ dependence (green line). 
The difference between the red and green lines for $\log a_* > -0.2$ ($a_* > 0.63$)
shows the strong influence of the magnetic field to the beam power for rapidly spinning BHs. In order of increasing $a_*$, the absolute errors in $\log L_{BH*}$ are 1.5, 0.65, 0.93, 0.49, 0.13, 0.16, 0.16, 0.28, 0.14, and 0.13.
\label{fig2}}
  \end{center}
  \vspace{-0.5cm}
\end{figure}

For the beam power, we follow \cite{liv99} and \cite{mei99,mei01}, and we find for two outflowing jets that 
\begin{eqnarray}
L_{BH}  & = & \frac{1}{2c}B_{eq}^2\ R_H^2\ \left(\Omega_H R_H\right)^2\nonumber \\
       & = & \frac{1}{32}B_{eq}^2\ R_S^2\ c \ a_*^2\ Q_*^2\nonumber \\
       & = & (6.6\times 10^{21}~{\rm W})\ {\dot m} \left(\frac{M}{M_\odot}\right) x^{-2.5} a_*^2\ Q_*^2\ ;
\label{Lbh1}
\end{eqnarray}
then we take the dimensionless beam power $L_{BH*}$ to be:
\begin{equation}
L_{BH*}\equiv \left(\frac{M}{M_\odot}\right) x^{-2.5}a_*^2\ Q_*^2\ ,
\label{Lbh2}
\end{equation}
in units of $(6.6\times 10^{21}~{\rm W}) {\dot m}$.
Fig.~\ref{fig2} shows $L_{BH*}$ from
eq.~(\ref{Lbh2}) vs. $a_*$ for the data listed in Table~\ref{table1}.
For comparison purposes, fiducial dashed lines show a pure $a_*^2$ dependence and an $M a_*^2$ dependence. In the region with
$a_*>0.63$ ($\log a_* > -0.2$), the actual data rise a lot higher than both dashed lines.
This indicates that the beam power does not vary as $a_*^2$ as is
commonly believed. The reason is the strong influence of the $x^{-2.5}$
term which in turn depends on $a_*$. This influence is most apparent
for high $a_*$ values, where the term increases the beam powers
substantially (by as much as a factor of 88 compared to the $a_*=0$ 
Schwarzschild case); 
and makes $L_{BH*}$ a much steeper function of $a_*$ than $a_*^2$.

As was observed in Fig.~\ref{fig1}, the $L_{BH*}$ values in Fig.~\ref{fig2}
also segregate
into two groups that show different slopes, one at moderate $a_*$ values and another at
high $a_*$ values. A linear regression of the lower 4 points 
with $a_*\leq 0.63$ shows a slope of 2.80 with a correlation
coefficient of $r^2=0.993$; a linear fit of the higher 6 points with
$a_*\geq 0.70$ shows a slope of 7.64 with $r^2=0.961$; and
a linear fit of all 10 points
shows a slope of 3.28 with $r^2=0.954$. Therefore, if all
the power in the observed transient (ballistic) jets is emitted by the BH alone,
the data should show a much steeper slope than the commonly
quoted slope of $d\log L_{BH*}/d\log a_* = 2$. In view of these
results, the current disagreement on whether such a slope of 2
can be deduced from the existing radio data \citep{fen10,nar12,rus13}
appears to be a moot issue. Furthermore, we have calculated that the best-fit line in the \cite{nar12} data (after resetting the mass of GRS 1915+105 to the value shown in Table~\ref{table1}) has a slope of $2.66\pm0.025(1\sigma)$ which lies within the error bar of our slope of $2.80\pm0.17(1\sigma)$ for moderate 
$a_* \leq 0.7$ where the 3 out of 4 radio sources lie. 
The next question is whether we can understand
the rather perplexing results of \cite{rus13} in the context of this investigation. We undertake this task in \S~\ref{new} below using also
Table~\ref{table2} in which we summarize the above results from linear
regressions and the regressions that we describe in subsequent sections.

\begin{table}
\caption{Summary of Results from Linear Regressions}
\label{table2}
\begin{tabular}{llll}
\hline
\multicolumn{4}{c}{Power from BH only}\\
\hline
 Model & \multicolumn{3}{c}{Best-Fit Slope (Correlation $r^2$)}\\
 & $a_*\leq 0.63$ pts. & $a_*\geq 0.7$ pts. & All 10 points \\ 
\hline
$L_{BH}\propto a_*^2$  & 2.80 (0.993) & 7.64 (0.961) & 3.28 (0.954) \\
$L_{BH}\propto a_*^4$  & 4.80 (0.998) & 9.64 (0.975) & 5.28 (0.982) \\
\hline 
\\
\multicolumn{4}{c}{Power from Disk and BH}\\
\hline
 Model & \multicolumn{3}{c}{Best-Fit Slope (Correlation $r^2$)}\\
  & $a_*\leq 0.63$ pts. & $a_*\geq 0.7$ pts. & All 10 points \\ 
\hline
$L_{BH}\propto a_*^2$  & 0.58 (0.903) & 4.52 (0.906) & 0.92 (0.708) \\
$L_{BH}\propto a_*^4$  & 2.58 (0.995) & 6.52 (0.953) & 2.92 (0.961) \\
\hline
\end{tabular}
\\
{Note.---Statistical $1\sigma$ errors in the quoted slopes are $\pm0.17$, $\pm0.77$, and $\pm0.26$ (BH only); and $\pm0.13$, $\pm0.73$, and $\pm0.21$ (Disk and BH); in each column respectively. A Maximum Likelihood method \citep{yo04} using the errors listed in Table~\ref{table1} produces similar results with larger errors in the left two columns, but deviates substantially in the third column where large weights are assigned exclusively to the high $a_*$ points that have small errors \citep[the algorithm is described by][]{thi11}. This indicates that a single straight line is not a good fit for all 10 points.
}
\end{table}

\section{Electromagnetic Output from the Black Hole and the Inner Disk}\label{total}

In this section, we add to the results of \S\ref{bhonly} the contribution
from the inner accretion disk. We assume a nonrelativistic accretion flow
\citep{liv99,mei99} but we have checked that the results
are not modified substantially in the case of a relativistic disk flow \citep{mei01}.
Again, the calculations incorporate
the same equipartition magnetic field $B_{eq}$   
(eq.~[\ref{ckb}]) for the region $R_H\leq R\leq R_I$
between the BH horizon and the ISCO.

The poloidal magnetic flux $\Psi$ now is
\begin{eqnarray}
\Psi & = & B_{eq}\left(\pi R_I^2\right)\nonumber \\
      & = & (6.2\times 10^{10}~{\rm Wb})\ {\dot m}^{0.5}
      \left(\frac{M}{M_\odot}\right)^{1.5} \left(4x^{0.75}\right),
\label{Psi1}
\end{eqnarray}
and we take the dimensionless flux $\Psi_*$ to be:
\begin{equation}
\Psi_*\equiv \left(\frac{M}{M_\odot}\right)^{1.5} \left(4x^{0.75}\right),
\label{Psi2}
\end{equation}
in units of $(6.2\times 10^{10}~{\rm Wb}) {\dot m}^{0.5}$.
The scaling of $\Psi_*$ is the same as in \S~\ref{bhonly} in order to facilitate
direct comparisons between eq.~(\ref{Psi2}) and eq.~(\ref{Psibh2}).
The parameter $x$ is again related to $a_*$ 
using the equations of \cite{bar72}.

\begin{figure}
\begin{center}
    \leavevmode
      \includegraphics[trim=0 1.3cm 0 1cm, clip, angle=0,width=9 cm]{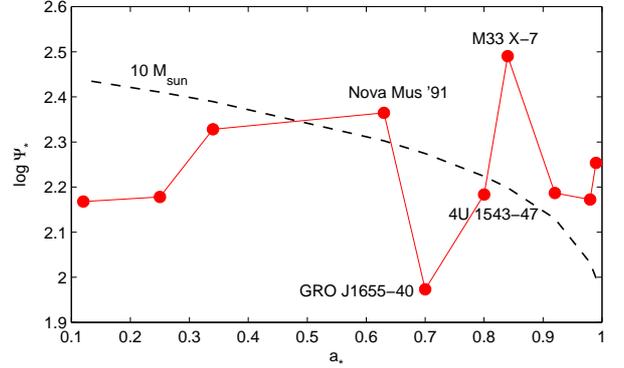}
\caption{Total poloidal magnetic flux $\Psi_{*}$ from
eq.~(\ref{Psi2}) vs. $a_*$ for the data listed in Table~\ref{table1}
and connected by straight line segments.
The dashed line shows the contribution from the disk and a BH with a fixed mass of 10$M_\odot$.
In order of increasing $a_*$, the absolute errors in $\log\Psi_{*}$ are 
0.01, 0.02, 0.02, 0.06, 0.01, 0.04, 0.03, 0.01, 0.08, and 0.00.
\label{fig3}}
  \end{center}
  \vspace{-0.5cm}
\end{figure}

We adopt again the data set for the 10 BHs listed in Table~\ref{table1},
although we have checked that the data from the Fe~K$\alpha$ method do not disagree with our determinations described below.
Fig.~\ref{fig3} shows $\Psi_{*}$ from
eq.~(\ref{Psi2}) vs. $a_*$ for the data listed in Table~\ref{table1}.
Again GRO J1655-40 appears to 
separate the fluxes into two groups, one with increased fluxes at 
moderate $a_*$ values and another at high
$a_*$ values. In this case, however, the fluxes at high $a_*$ values are not dominant (compare Fig.~\ref{fig3} to Fig.~\ref{fig1} for $a_*\geq 0.8$).
But it is hard to tell whether this division is significant with so few data available.
We note that $\Psi_{*}$ decreases with $a_*$ for a fixed mass such as the 10$M_\odot$ BH
plotted as a dashed line in Fig.~\ref{fig3}. This is the result of the decrease of the
disk term $x^{0.75}$ with $a_*$ in eq.~(\ref{Psi2}) and it occurs despite the gradual increase of the contribution from around the BH horizon. Therefore, we have theoretical evidence that the magnetic flux becomes subdued with increasing $a_*$ at high
$a_* \geq 0.8$ values when it is dominated by the field lines that thread the inner accretion disk.

\begin{figure}
\begin{center}
    \leavevmode
      \includegraphics[trim=0 1.3cm 0 1cm, clip, angle=0,width=9 cm]{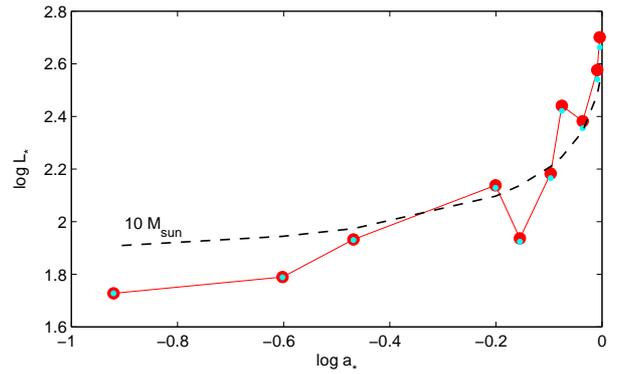}
\caption{Total beam power $L_{*}$ from
eq.~(\ref{L2}) vs. $a_*$ for the data listed in Table~\ref{table1}
and connected by straight line segments.
The cyan dots show the disk's contribution to the total power for each object in Table~\ref{table1}. The dashed line shows the combined contribution from the disk and a BH with a fixed mass of 10$M_\odot$.
In order of increasing $a_*$, the absolute errors in $\log L_{*}$ are 
0.41, 0.19, 0.24, 0.15, 0.16, 0.17, 0.16, 0.21, 0.09, and 0.05.
\label{fig4}}
  \end{center}
  \vspace{-0.5cm}
\end{figure}

For the combined beam power $L$, we use $v=\sqrt{GM/R_I}$ for the orbital speed of the plasma at the ISCO and we find for two outflowing jets that
\begin{eqnarray}
L    & = & L_{Disk} + L_{BH}\nonumber \\
     & = & B_{eq}^2\ R_I^2\ v  +  
                 \frac{1}{32}B_{eq}^2\ R_S^2\ c\  a_*^2\ Q_*^2\nonumber \\
      & = & (1.5\times 10^{23}~{\rm W})\ {\dot m} \left(\frac{M}{M_\odot}\right) x^{-1}\nonumber \\
      &    &  +\ (6.6\times 10^{21}~{\rm W})\  {\dot m} \left(\frac{M}{M_\odot}\right) x^{-2.5} a_*^2\ Q_*^2 \ ;
\label{L1}
\end{eqnarray}
then we take the dimensionless beam power $L_{*}$ to be:
\begin{equation}
L_{*}\equiv \left(\frac{M}{M_\odot}\right) \left(16\sqrt{2}\ x^{-1} + x^{-2.5}a_*^2\ Q_*^2\right) ,
\label{L2}
\end{equation}
in units of $(6.6\times 10^{21}~{\rm W})  {\dot m}$.
Fig.~\ref{fig4} shows $L_{*}$ from
eq.~(\ref{L2}) vs. $a_*$ for the data listed in Table~\ref{table1}.
The $L_{*}$ values in Fig.~\ref{fig4} can be separated 
into two groups that show different slopes, one at moderate $a_*$ values and another at
high $a_*$ values. A linear regression of the lower 4 points 
with $a_*\leq 0.63$ shows a slope of 0.58 with a correlation
coefficient of $r^2=0.903$; a linear fit of the higher 6 points with
$a_*\geq 0.70$ shows a slope of 4.52 with $r^2=0.906$; and
a linear fit of all 10 points
shows a slope of 0.92 with $r^2=0.708$. The 
importance of the disk's contribution to the emitted power is
apparent in these results.
The power from the BH is too small to support the notion
that a slope of 2 is significant. Therefore, if the inner accretion
disk contributes significantly to compact jet-like poloidal outflows, we expect that
the data should show a shallow slope of order 
$d\log L_{*}/d\log a_* \approx 0.6-0.9$, at least for $a_* < 0.8$
if not for the entire $a_*$ range. In principle, this result does not disagree with the
analysis of the observations of continuous compact jets \citep{fen10}. Although the
data illustrated in Fig.~4 of this study do not show an obvious linear correlation, 
any line that may be fitted will certainly have a very shallow slope. This indicates
that the continuous compact jets seen in the hard states of these objects originate mostly from the
inner accretion disks and not from the BHs 
\citep[see also][]{kyl12}.

\section{A New Estimate of the Magnetic Field Around the Black Hole}\label{new}

The equations that we have used in the previous sections to estimate the beam power
from the BH and the inner nonrelativistic disk are consistent with the jet power expected
to be emitted by the BZ mechanism \citep{bla77} and the BP mechanism \citep{bla82,con94}, respectively \citep{liv99,mei99}.
The only novel element in our calculations has been so far the adoption of the equipartition poloidal
magnetic field from the work of \cite{ck98} on the PRCB (eq.~[\ref{ckb}]). This equation, along
with the location of the ISCO (eq.~[\ref{x_par}]), has introduced the nonlinear dependence of the beam power
on the spin parameter $a_*$ as seen in Figs.~\ref{fig2} and~\ref{fig4}. In the case of
the BH power, the explicit dependence of $L_{BH}$ on $B_{eq}^2 a_*^2$ that is derived
in the BZ mechanism can be seen in eq.~(\ref{Lbh1}), and it is the magnetic field
that introduces a strong nonlinearity (via the $x^{-2.5}$ term that varies by
a factor of 88 from $a_*=0$ to 1) that
spoils the pure $a_*^2$ dependence in eq.~(\ref{Lbh2})
(the term $Q_*^2$ is not as important as it varies by only a factor of 4).
However, these classical equations may not be correct because they assume that the accumulated magnetic field can reach equipartition undisturbed. 

\begin{figure}
\begin{center}
    \leavevmode
      \includegraphics[trim=0 1.3cm 0 1cm, clip, angle=0,width=9 cm]{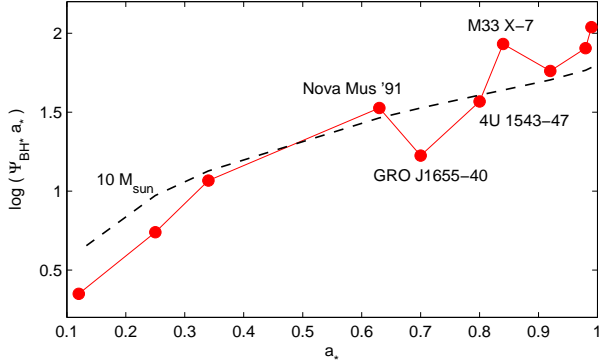}
\caption{As in Fig.~\ref{fig1}, but an additional factor of $a_*$ has been inserted
into the flux $\Psi_{BH*}$ in eq.~(\ref{Psibh2}).
The dashed line shows a BH with a fixed mass of 10$M_\odot$.
In order of increasing $a_*$, the absolute errors in $\log(\Psi_{BH*}a_*)$ are 0.76, 0.35, 0.47, 0.29, 0.07, 0.11, 0.10, 0.15, 0.12, and 0.07.
\label{fig5}}
  \end{center}
  \vspace{-0.5cm}
\end{figure}

\begin{figure}
\begin{center}
    \leavevmode
      \includegraphics[trim=0 1.3cm 0 1cm, clip, angle=0,width=9 cm]{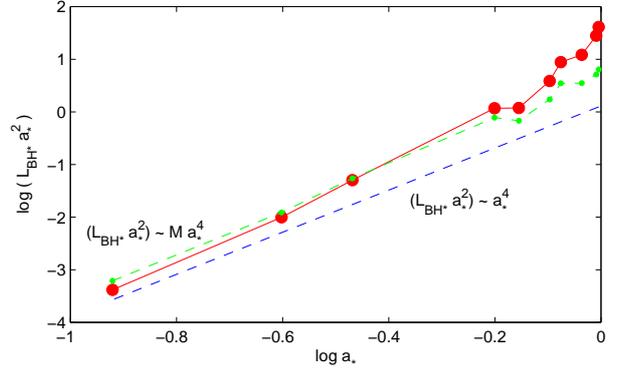}
\caption{As in Fig.~\ref{fig2}, but an additional factor of $a_*^2$ has been inserted
into the power $L_{BH*}$ in eq.~(\ref{Lbh2}).
The dashed lines show a pure $a_*^4$ dependence (blue line)
and an $M a_*^4$ dependence (green line). 
The difference between the red and green lines for $\log a_* > -0.2$ ($a_* > 0.63$)
shows the strong influence of the magnetic field to the beam power for rapidly spinning BHs. In order of increasing $a_*$, the absolute errors in $\log (L_{BH*}a_*^2)$ are 2.9, 1.2, 1.6, 0.73, 0.19, 0.22, 0.21, 0.35, 0.14, and 0.14.
\label{fig6}}
  \end{center}
  \vspace{-0.5cm}
\end{figure}

The recent work of \cite{con16} indicates that the region
$R_H\leq R\leq R_I$ cannot support an equipartition
magnetic field unless the BH is maximally rotating ($a_*=1$). 
The origin of the problem is the magnetic
Rayleigh-Taylor (MRT) instability that allows for the poloidal field to escape by slipping in the azimuthal direction. On the other hand, a fast rotation of the spacetime works toward limiting the instability.
A simplified version of the
results of \cite{con16}
is that the
maximum magnetic field that can thread the BH horizon 
and the inner disk is proportional to the BH spin itself, that is
\begin{equation}
B_{max} \approx B_{eq} \ a_* ,
\label{newb}
\end{equation}
to within factors of order unity. This new result challenges the fundamentals of jet emission from BH binaries (without actually negating the BZ mechanism) and forces us to reconsider the classical picture that was analyzed in \S\S\ref{bhonly} and~\ref{total} above. 
In the following subsections, we adopt eq.~(\ref{newb}) to provide a description of the maximum magnetic field
around the BH and the inner disk, and we repeat the analyses of \S\S\ref{bhonly} and~\ref{total}.

\subsection{Black Hole}\label{new_bh}

We insert a factor of $a_*$ into $\Psi_{BH*}$ in eq.~(\ref{Psibh2})
and a factor of $a_*^2$ into $L_{BH*}$ in eq.~(\ref{Lbh2}). 
Figs.~\ref{fig5} and~\ref{fig6} are analogous to Figs.~\ref{fig1} and~\ref{fig2} and they show the resulting changes in
the physical quantities involved in the emission from the BH only.

The new dependence of $B_{max}$ on the spin parameter has steepened dramatically the slopes of the curves depicted in Figs.~\ref{fig5} and \ref{fig6}. Again, we can separate the results into two groups that show different slopes, one at moderate $a_*$ values and another at high $a_*$ values. By repeating the linear regressions outlined in
\S\ref{bhonly}, we find that the new slopes of the best-fit lines to the beam powers now stand higher by $+2$ while the correlation coefficients are also higher (Table~\ref{table2}). Therefore, if all
the power in the observed transient jets is emitted by the BH alone;
and if the MRT instability limits the magnetic field around the BH
to obey eq.~(\ref{newb}); then the radio observations
should show dramatically steeper slopes ($\approx$ 5-10)
than the commonly
quoted slope of $d\log L_{BH*}/d\log a_* = 2$. 

This model may help explain some of the perplexing
results obtained by \cite{rus13} from radio observations of jets
in a large sample of BH binaries. In particular, their Figs.~1c
and~1d show clearly two groups of points; one at low/moderate
$a_*\leq 0.6$ values with slopes of about 2.5-3 (but still not close to our 4.80);
and another at high $a_*\geq 0.63$ values with slopes of about 8-15 
(roughly comparable to our 9.64). Given the approximations involved
in the radio jet data analysis and in our analysis, we find the rough agreement
between these results satisfactory. But we must note that the results 
of \cite{nar12} and \cite{rus13} for $a_*\leq 0.63$ seem to support the conventional $L_{BH*}\propto a_*^2$ model of \S~\ref{bhonly} 
(see Table~\ref{table2}) and not the new $a_*^4$ model (but see also \S~\ref{new_disk} and the discussion at the end of \S~\ref{conc} below).

\begin{figure}
\begin{center}
    \leavevmode
      \includegraphics[trim=0 1.3cm 0 1cm, clip, angle=0,width=9 cm]{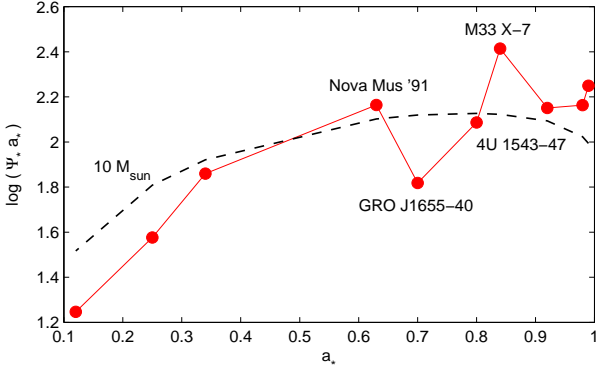}
\caption{As in Fig.~\ref{fig3}, but an additional factor of $a_*$ has been inserted
into the flux $\Psi_{*}$ in eq.~(\ref{Psi2}).
The dashed line shows the contribution from the disk and a BH with a fixed mass of 10$M_\odot$.
In order of increasing $a_*$, the absolute errors in $\log(\Psi_{*}a_*)$ are 0.68, 0.28, 0.33, 0.18, 0.04, 0.07, 0.05, 0.05, 0.08, and 0.01.
\label{fig7}}
  \end{center}
  \vspace{-0.5cm}
\end{figure}

\begin{figure}
\begin{center}
    \leavevmode
      \includegraphics[trim=0 1.3cm 0 1cm, clip, angle=0,width=9 cm]{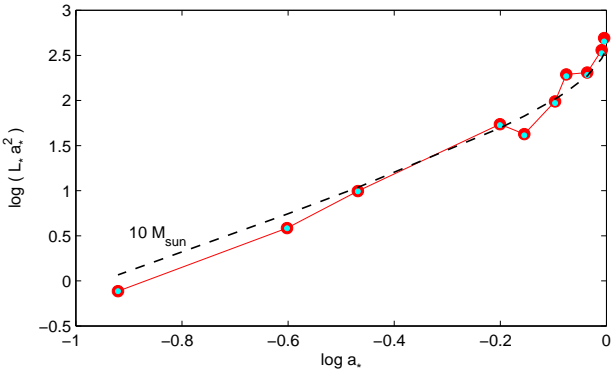}
\caption{
As in Fig.~\ref{fig4}, but an additional factor of $a_*^2$ has been inserted
into the power $L_{*}$ in eq.~(\ref{L2}).
The cyan dots show the disk's contribution to the total power for each object in Table~\ref{table1}. The dashed line shows the combined contribution from the disk and a BH with a fixed mass of 10$M_\odot$.
In order of increasing $a_*$, the absolute errors in $\log (L_{*}a_*^2)$ are 1.8, 0.71, 0.95, 0.40, 0.22, 0.22, 0.21, 0.28, 0.10, and 0.06.
\label{fig8}}
  \end{center}
  \vspace{-0.5cm}
\end{figure}

\subsection{Accretion Disk and Black Hole}\label{new_disk}

We insert a factor of $a_*$ into $\Psi_{*}$ in eq.~(\ref{Psi2})
and a factor of $a_*^2$ into $L_{*}$ in eq.~(\ref{L2}). 
Figs.~\ref{fig7} and~\ref{fig8} are analogous to Figs.~\ref{fig3} and~\ref{fig4} and they show the resulting changes in
the physical quantities involved in the combined emission from the
accretion disk and the BH.

By repeating the linear regressions outlined in
\S\ref{total}, we find that the new slopes of the best-fit lines to the beam powers now stand higher by $+2$ while the correlation coefficients are  substantially higher (compare the two bottom rows in Table~\ref{table2}). 
A surprising element is the linear fit
to the lower 4 points ($a_*\leq 0.63$) that produces a line with slope 2.58 and an excellent correlation ($r^2=0.995$). This slope
is comparable to that of the $a_*^2$ model with a BH only
(2.80; Table~\ref{table2}) and to the slopes seen in the 
\cite{nar12} and \cite{rus13} data. As a result, the plot thickens since the emission from dominant $a_*^4$ disks around moderately
spinning BHs appears to have the same signature slope as the power output
from $a_*^2$ BHs surrounded by dormant disks. 
Theoretical considerations alone cannot tell the difference between these two cases and observations will need to turn to $a_*\geq 0.7$
systems in which active $a_*^4$ disks are incapable of producing slopes as high as those found for conventional $a_*^2$ BHs
(see the entries for $a_*\geq 0.7$ in Table~\ref{table2}).
The $a_*\geq 0.7$ regime appears to also be appropriate for distinguishing between the two $a_*$ models from observations of transient (ballistic) jets. If such jets are produced by rapidly-spinning BHs, then Table~\ref{table2} indicates that the 
$d\log L_{BH*}/d\log a_*$
slope predicted by the $a_*^4$ model is considerably steeper
(9.64 as opposed to 7.64 for the $a_*^2$ model).

\section{Summary and Discussion}\label{conc}

In this investigation, we have calculated the physical parameters associated with jet emission from BH binaries, the poloidal magnetic flux and the beam (jet) power. We were aided by
an homogeneous data set of BH masses and spins that were determined by the CF method
(Table~\ref{table1}).
The results do not appear to be modified in a substantial way when we examine data from the 
Fe K$\alpha$ method or when we adopt relativistic accretion disks, and we opted to not show another set of results for the sake of clarity. Our results can be summarized as follows:
\begin{itemize}
\item[1.] The (unobservable) poloidal magnetic flux indicates that the objects may be segregated into two broad groups, one with low/moderate spins, and another with very high spins. This effect is apparent in the classical model \citep{bla77,bla82} calculations depicted in Figs.~\ref{fig1} and~\ref{fig3}, but it is harder to detect in the new model of jet emission \citep{con16} shown in Figs.~\ref{fig5} and~\ref{fig7}. The two separated groups are created by the particular dependence
of the fluxes on the BH masses listed in Table~\ref{table1}. A similar effect is seen
in the Fe~K$\alpha$ data as well (see footnote~\ref{ft2}).
\item[2.] The beam power from the vicinity of the BH does not scale with spin as $a_*^2$ (Fig.~\ref{fig2}) because both the location of the ISCO and the supported magnetic field depend nonlinearly on $a_*$ \citep{bar72,ck98,con12}. Linear fits to the data indicate that the actual scaling is steeper and closer to 2.8-3.3 for most of the range of $a_*$ values and close to 7.6 for high $a_*$ values (\S~\ref{bhonly}, Table~\ref{table2}). In view of these results, the
controversy that has developed about the $a_*^2$ scaling \citep{fen10,nar12,rus13} now 
appears to be moot. Interestingly, both sets of results
can now be interpreted in a consistent manner 
(see the discussions at the end of \S~\ref{bhonly} and at the end of \S~\ref{new_bh}). 
\item[3.] When the contribution of the inner accretion disk is included, the disk power output dominates over the output produced near the BH (\S~\ref{total}, Figs.~\ref{fig3} and~\ref{fig4}). Then, linear fits to the data listed in Table~\ref{table2} indicate that the actual scaling of the beam power is quite shallow ($d\log L_{*}/d\log a_* \approx 0.6-0.9$ 
at least over the interval of $0\leq a_* < 0.8$).
\item[4.] A new model of jet power, the $a_*^4$ model  \citep{con16}, was outlined in \S~\ref{new}; as shown in Table~\ref{table2}, it predicts higher slopes (by an additional $+2$) in various sections of the data relative to the slopes determined for the conventional $a_*^2$-power models. 
Interestingly, very steep slopes were found also by \cite{rus13} 
in the high $a_*$ regime when they analyzed the available radio observations of BH transient (ballistic) jets. 
\item[5.] The current observational data \citep{nar12,rus13}
for systems with moderate $a_*\leq 0.63$ values
do not let us distinguish between the two $a_*$ models or between
emission from a dominant $a_*^2$ BH versus from an $a_*^4$ BH
with a dominant inner accretion disk. The results discussed in \S~\ref{new_disk} indicate that observations of jets from 
rapidly-spinning BHs ($a_*\geq 0.7$) may help us resolve both of these issues in the future, but it will not be easy: in the former case, the difference between slopes is only $+2$ in all cases and in all $a_*$ intervals; and in the latter case, the slopes differ only by about
$+1$ (a slope of 7.64 vs. 6.52 in Table~\ref{table2}).
\end{itemize}

The steep dependence of the power output on $a_*$ seen
in Table~\ref{table2} for $a_*\geq 0.7$ has also been seen
in simulated relativistic models of BHs accreting from surrounding ADAF-type disks. For BH jets and for $a_* > 0.5$, \cite{mck05} found that the best-fit slope was 5 whereas \cite{tch10} found values as high as 6 for the thickest of their ADAF models with $a_* > 0.7$. These results seem to point in the right direction, but they still lie below our most conservative slope (7.64; Table~\ref{table2}).
Furthermore, \cite{mck05} obtained a fit to the total power output in these models and a slope of 4. This result points again in the right direction: Table~\ref{table2} shows that the slope of the total power decreases by about 3 units compared to the pure BH power.
In this case, the slope of 4 of \cite{mck05} appears to compare favorably with our slope of 4.52 derived from the $a_*^2$ BH model with a dominant accretion disk.

On the other hand, the steep slope of 4.80 found in the new $a_*^4$ BH model (\S~\ref{new_bh}) for $a_*\leq 0.63$ is not borne out by
the analyses of radio observations of \cite{nar12} and \cite{rus13}
that indicate typical slopes of about 2.5-3. These studies show that,
in low $a_*$ objects, the ballistic jet power does not decrease with decreasing $a_*$ as strongly as implied by the $a_*^4$ BH model.
We have tried to understand this outcome as follows: In the $a_*^4$ BH model \citep{con16}, the MRT instability limits the accumulated magnetic field $B_{max}$ to a value well below the equipartition value $B_{eq}$ in low $a_*$ objects (eq.~[\ref{newb}]), but this will not be reflected in the data if the magnetic field is continually produced by the PRCB and the accretion occurs on dynamical (free-fall) timescales.
The MRT instability also acts dynamically to remove the accumulated magnetic flux near the BH. 
One may then argue that for half the time, the flux is brought into the vicinity of the BH from the ISCO; and for the other half, a fraction of the flux, proportional to $(1-a_*)B_{eq}$, is removed from the same area. As a result, the average (integrated over time and divided by the total time interval) accumulated magnetic field will not decrease to 
that given by eq.~(\ref{newb}) which is stable against the MRT instability; but it will remain instead near the time-integrated average value that is proportional to $(1+a_*)B_{eq}/2$. This estimate yields the expected behavior for $a_*\to 1$ ($B_{max}\to B_{eq}$); but for $a_*\to 0$, it shows that a substantial fraction of the equipartition magnetic field ($B_{eq}/2$) will remain near the BH horizon.
In this case (of ``the $(1+a_*)^2a_*^2/4$ model''), a linear regression to the objects of Table~\ref{table1} with
$a_*\leq 0.63$ shows a slope of 3.25 ($r^2=0.990$) as opposed to the slope of 4.80 listed in the second row of Table~\ref{table2}.

\section*{Acknowledgments}
We thank the referee whose comments led to a thorough analysis of the error bars in our calculations.
DMC and SGTL were supported in part by NASA grant NNX14-AF77G.
DK was supported by a NASA ADAP grant.
JFS was supported by NASA Einstein Fellowship grant PF5-160144.

\label{lastpage}

\end{document}